\documentclass[12pt]{irsreport}
\usepackage{epsfig}

\title{IRS-TR 12003:  Constructing Low-Resolution Truth Spectra
of the Standard Stars HR~6348 and HD~173511}

\author{
G.C. Sloan (1) ~\& D. Ludovici (2) \thanks{ (1) Infrared Spectrograph 
Science Center, Cornell University, (2) Department of Physics and
Astronomy, University of Iowa; NSF REU Research Assistant, Astronomy 
Department, Cornell University} }

\date{19 December, 2012}

\runninghead{IRS-TR 12003 - Low-Resolution Truth Spectra}

\begin{document}

\maketitle

\begin{abstract}

This report describes the generation of fully calibrated 
spectra of the K giants HR~6348 and HD~173511 from data 
obtained with the low-resolution modules of the Infrared 
Spectrograph (IRS), with an emphasis on the spectra from the 
Long-Low (LL) module.  The spectra were calibrated using 
Kurucz models and IRS observations of the A dwarfs 
$\alpha$~Lac and $\delta$~UMi.  The calibration process 
required mitigation for fringing in the first-order LL 
spectrum and a faint red excess in $\alpha$~Lac which may 
arise from a low-contrast debris disk.  The final calibrated 
spectrum of HR~6348 has a spectroscopic fidelity of 0.5\% or 
better below 29~\mum, with an uncertainty increasing to 
$\sim$1\% at 33--37~\mum.  The final calibrated spectrum of 
HD~173511 has a spectroscopic fidelity of $\sim$0.5\% at all 
wavelengths below 35.8~\mum.

\end{abstract}

\section{Introduction} 

IRS-TR 12002 (Sloan \& Ludovici 2012b) described the 
generation of a ``truth'' spectrum of the K giant HR~6348 for 
use with Short-Low (SL) module of the IRS.  Truth spectra are 
fully calibrated spectra of a standard star which can in turn
be used to calibrate other spectra.  The approach was to use 
SL observations of HR~6348, combined with observations and 
Kurucz models of the A dwarfs $\alpha$~Lac and $\delta$~UMi.  
This report describes the procedure for the Long-Low (LL) 
module and the assembly of full low-resolution spectra using 
both SL and LL.  The resulting spectra for HR~6348 and 
HD~173511 can be used as truth spectra for the calibration of 
all LL observations.  The method for LL is generally similar 
to that for SL.

The detector settings for LL were altered at the beginning
of IRS Campaign 45 (29 October, 2007), requiring different
calibrations for data taken through Campaign 44 and data
taken from Campaign 45 to 61, when the cryogenic {\it
Spitzer} mission ended.

\section{Procedure} 

As described in IRS-TR 12002, we begin the calibration
process with a Kurucz model of an early A dwarf supplied to
the IRS Team by M.\ Cohen, shifted to the wavelength grids 
and resolutions of SL and LL, and scaled photometrically to 
match the Red Peak-Up (PU) photometry of $\alpha$~Lac and 
$\delta$~UMi.

We first generated a calibrated LL spectrum of HR~6348
using data from Campaign P in the Science Verification phase 
and Campaigns 1--44 during normal science operations.  We 
then used this spectrum to calibrate all of the other data 
from the same time period and from those data constructed a 
calibrated spectrum of HD~173511.  This star has a spectral 
class of K5 III; its later spectral class leads to a deeper 
SiO absorption band at 8~\mum\ compared to HR~6348, and for 
this reason we have avoided its use to calibrate SL.  It is 
still a good standard for LL, and because of the limited 
calibration data available from the change in detector 
settings for LL in Campaign 45 to the end of the cold {\it 
Spitzer} mission, both stars are needed to produce a spectral 
correction with a high signal/noise ratio.

The procedure follows the general outline used in SL, where
we combined the calibrations of HR~6348 using $\alpha$~Lac
and $\delta$~UMi order by order, assembled the orders, and
then mitigated for any remaining detectable artifacts.  
However, the detailed procedure differs in several ways 
from SL, due to differences in instrumental performance and 
stellar emission in the two wavelength regimes.

We used all of the HR~6348 spectra observed from IRS Campaign
P through Campaign 44.  For SL, we excluded poorly pointed 
data, but for LL, the throughput is largely unchanged from 
one pointing to the next because the pointing errors are 
small compared to the larger size of the LL slit (10'').

We also changed how we combine the spectra from the two nod
positions.  In SL, we carefully compared the separate 
spectra, avoiding structure in one nod position not seen in 
the other.  In LL, however, the spectra from the two nods 
show similar structure on fine wavelength scales, and we have 
simply averaged them.  

In SL, we used a spline-smoothing algorithm to force the 
spectra calibrated with $\alpha$~Lac and $\delta$~UMi to a 
similar averaged shape, and again carefully avoided structure 
in one calibration not seen in the other.  Here again the LL 
procedure differs.  Both calibrations showed similar degrees 
of structure, so both were used at all wavelengths.  However, 
the overall shapes of the two spectra differ, with the 
spectrum of HR~6348 dropping more steeply to longer 
wavelengths when calibrated with $\alpha$~Lac than with 
$\delta$~UMi.  From 15 to 37~\mum, the two calibrations have 
a 3\% difference in slope.

\begin{figure} 
  \begin{center}
     \epsfig{file=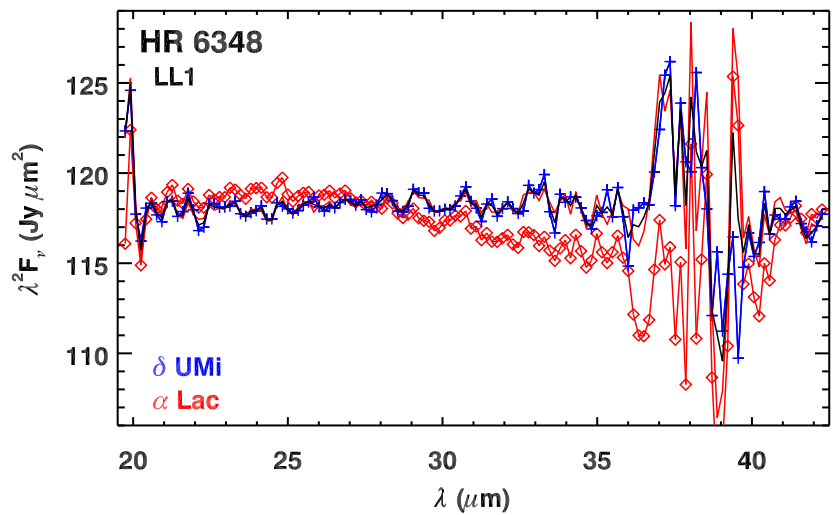, width=11cm}
  \end{center}
\caption{
---The construction of the LL1 spectrum of HR~6348.
The colored traces with symbols shows the input
spectra of HR~6348, as calibrated with $\delta$~UMi (blue)
and $\alpha$~Lac (red).  The two spectra show a considerable
difference in overall slope, which we attribute to a 
low-contrast red excess in $\alpha$~Lac, possibly from a
debris disk.  The red line without symbols show the
calibration based on $\alpha$~Lac spline-shifted to align 
with the calibration with $\delta$~UMi (which is unchanged by 
this step).  The final combined spectrum from the two 
calibrations appears in black.  The signal/noise ratio 
deteriorates rapidly past 36~\mum.}
\end{figure}

Figure~1 shows the spectrum of HR~6348 as calibrated with the
two A dwarfs in LL Order 1 (LL1).  The calibration based on 
$\alpha$~Lac actually falls more steeply than a Rayleigh-Jeans
tail (which would be horizontal in the units plotted) and is
not physically realistic.  The most likely explanation is that
the IRS spectrum of $\alpha$~Lac includes a red excess, which 
while low-contrast, appears to be real.  It probably arises
from a debris disk.  Consequently, we use the spline 
averaging in all three LL orders to force the overall shape 
of the calibration based on $\alpha$~Lac to align 
with the calibration based on $\delta$~UMi while preserving 
the spectral structure on finer wavelength scales.

\section{Long-Low Order 1} 

Figure~1 illustrates the construction of the LL1 spectrum of
HR~6348.  For wavelengths between 20 and 36~\mum, the 
structure in the spectrum generally deviates by less than 
$\pm$1\% from a Rayleigh-Jeans tail.  Moving past 36~\mum, 
the quality of the data grow rapidly worse with increasing
wavelength.  The apparent improvement at 40~\mum\ is 
misleading.  Beyond 40~\mum, a second-order spectrum
overlaps the first order, leading to contamination of the 
spectrum with emission from wavelengths of $\sim$20~\mum.  

\section{Long-Low Order 2} 

\begin{figure} 
  \begin{center}
     \epsfig{file=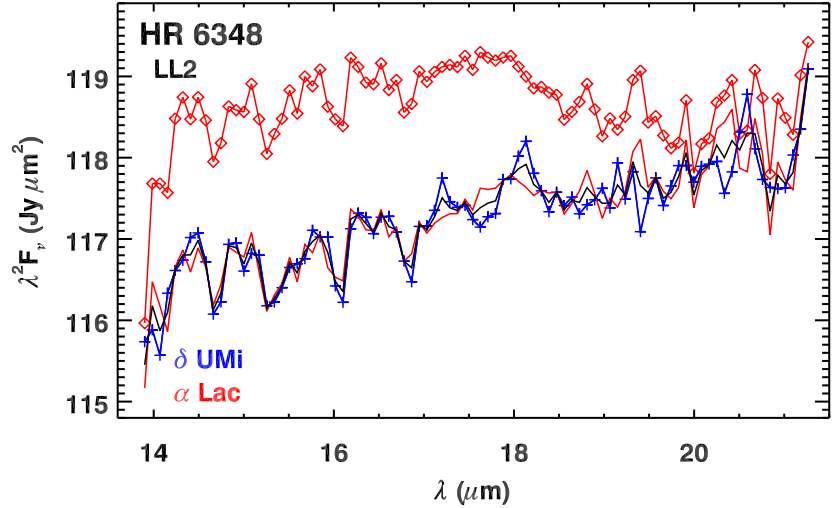, width=11cm}
  \end{center}
\caption{
---The contruction of the LL2 spectrum of HR~6348.  As with
LL1, the calibration based on $\alpha$~Lac is bluer due to 
the red excess in $\alpha$~Lac itself.  Symbols and colors 
are as defined in Fig.~1.  The spectral structure from 
$\sim$14 to 17~\mum\ is due to OH absorption in HR~6348 and 
is not an artifact.}
\end{figure}

Figure~2 shows the construction of the LL2 spectrum
of HR~6348.  The procedure is identical to LL1, with the
calibration using $\alpha$~Lac forced to the general shape
of the $\delta$ UMi calibration to compensate for the red
excess in the former.  Excluding the endpoints of the 
spectrum, all of the structure is less than 1\% of the 
continuum, peak-to-peak.  Below $\sim$17~\mum, the structure 
is intrinsic to HR~6348 and is due to OH band absorption.

\section{Long-Low Bonus Order} 

\begin{figure} 
  \begin{center}
     \epsfig{file=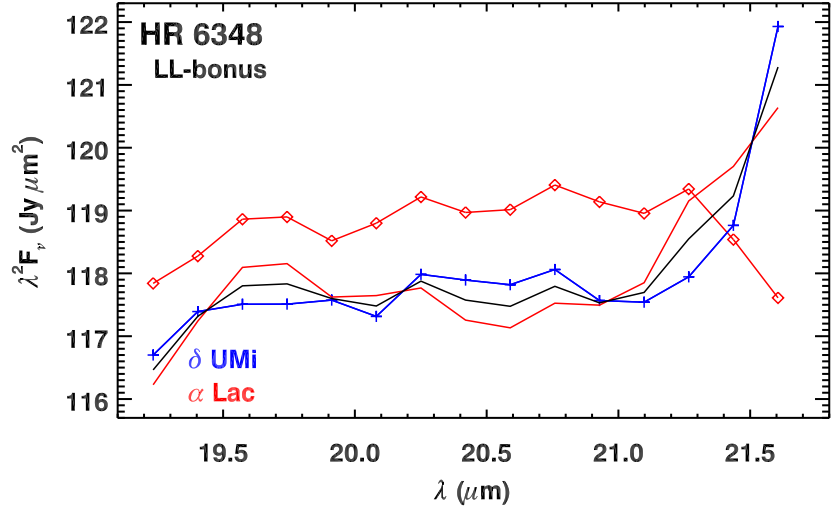, width=11cm}
  \end{center}
\caption{
---The contruction of the LL-bonus spectrum of HR~6348.  
Colors and symbols are as defined in Fig.~1.}
\end{figure}

Figure~3 illustrates the construction of the spectrum of 
HR~6348 in the LL-bonus order.  The procedure is identical
to that used in the other two orders.

\begin{figure} 
  \begin{center}
     \epsfig{file=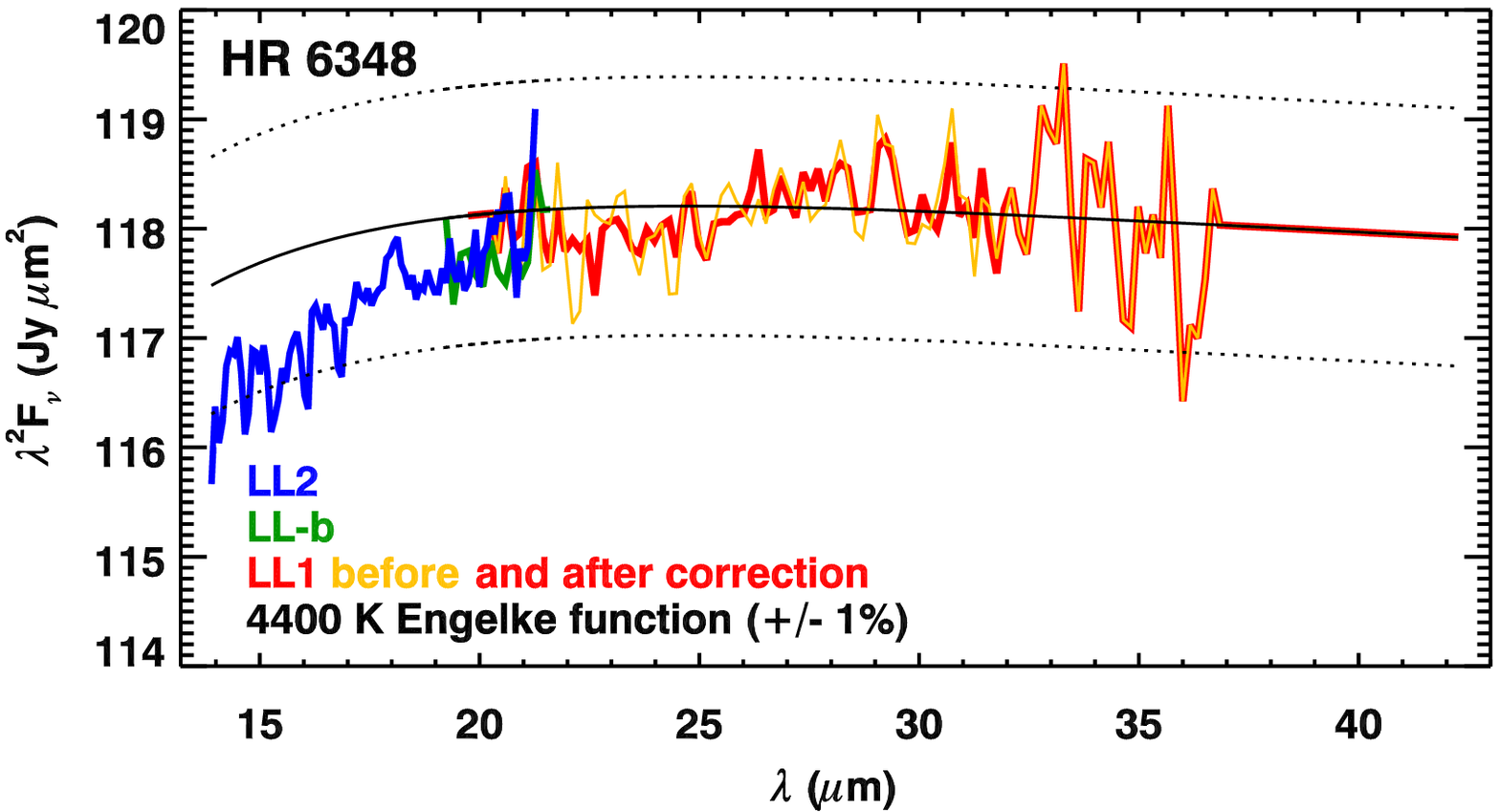, width=13.8cm}
  \end{center}
\caption{
---Completing the LL spectrum of HR~6348.  The orders are 
coded by color, with the LL1 data before the fringing 
correction in orange and after in red.  A 4400~K Engelke 
function fitted to the long-wavelength data is plotted in 
black, through the LL1 data and also shifted by $\pm$1\%.  
The deviations from an Engelke function in LL2 and LL-bonus 
appear to be real.  The spectroscopic fidelity of the 
spectrum is better than $\sim$0.5\% for $\lambda < $ 
28~\mum, and generally better than 1.0\% out to 37~\mum.  
Beyond 37~\mum, the data are too noisy to be of value.}
\end{figure}

\bigskip
\begin{center}
\begin{tabular}{lcc} 
\multicolumn{3}{c}{\bf Table 1---Wavelength Ranges} \\
\hline
{\bf Spectral segment} & {\bf Full range (\mum)} & {\bf Useful range (\mum)} \\
\hline
LL2      & 13.90--21.27  & 13.98--20.54 \\
LL-bonus & 19.23--21.61  & 19.40--21.10 \\
LL1      & 19.74--42.27  & 20.59--36.85 \\
\hline
\end{tabular}
\end{center}
\bigskip

Figure~4 presents the full LL spectrum of HR~6348 after
mitigating for the remaining problems.  A 4400 K Engelke
function fitted to the range of valid data in LL1 serves as 
a good reference (Engelke 1992).  We have also used it to 
replace invalid data at the ends of some of the segments.  
In LL1, this includes all data past 37~\mum, where the 
combination of dropping responsivity in the detector and the 
1/$\lambda^2$ dependence of the emission from the star leaves
the S/N of our coadded spectrum too low to be trustworthy.  
We have also used the fitted Engelke function to replace 
other data outside the useful range of the LL orders where 
those data are unreliable.  These ranges include the first 
four wavelength elements of LL1 (19.74--20.25~\mum), and for 
the LL-bonus order, the first element (19.23~\mum) and the 
last two (21.43--21.61~\mum).  We have left the remaining 
data outside the useful ranges alone because they are close 
enough to the expected values that replacement is 
unwarranted.

Figure~4 plots the LL1 data before and after a correction
for residual fringing.  This fringing arises from 
interference caused by partial delamination of the LL1 filter 
prior to launch.  It generally cancels out when a spectral 
correction is applied, but residuals can remain due to 
differences in pointing between a science target and the 
suite of calibrators.  In HR~6348, this residual fringing 
survives at a level of $\sim$0.6\%.  Because the OH band 
structure apparent in LL2 continues into LL1 with a 
periodicity similar to the fringing, we cannot use the 
HR~6348 data to determine a fringing solution.  Instead, we 
have used LL1 spectra of two red standards, Mrk~231 and 
IRAS~07598+6508.  For these sources, we fitted the continuum 
with a sixth-order polynomial and assumed that any deviations 
were due to fringing.  We reduced the peak-to-peak amplitude 
of the resulting correction by a factor of two before 
applying it to HR~6348 to account for the stronger fringing 
observed in the red standards.  We also limited our 
correction to wavelengths below 32~\mum\ because at longer 
wavelengths, the correction appeared to add more structure 
and/or noise to the spectrum than it removed.

\section{Quality Assessment} 

The Engelke function in Fig.~4 provides a means of assessing 
the quality of the data in LL1, where the largest diversions 
are due to noise.  The decreasing S/N from 21 to 37~\mum\ is 
readily apparent.  Below $\sim$29~\mum, the data are 
generally within $\pm$0.5\% of the Engelke function, and 
beyond $\sim$29~\mum, the bounds are about $\pm$1.0\%.  
As mentioned above, the noise beyond 37~\mum\ has led us 
to use an Engelke function as the best means of 
estimating the true spectrum of HR~6348 despite its 
limitation of including no spectral structure.

The LL-bonus spectrum is within 0.5\% of the Engelke 
function, although at these wavelengths, the actual spectrum 
of HR~6348 appears to be deviating from the shape of the 
Engelke function.  This trend continues into LL2, and because 
it reflects the differences in the shapes of the spectra of 
HR~6348 and the A dwarf $\delta$~UMi, it indicates that the 
Engelke function does not accurately portray the shape of an 
early K giant over such a broad wavelength range.  Most of 
the structure in the spectrum of LL2 arises from OH molecules 
and is probably real.  The exception is the apparent emission 
feature at 18~\mum, which is likely to be an artifact.  Even 
then, it is smaller than 0.5\%.  We conclude that below 
$\sim$29~\mum, the spectroscopic (point-to-point) fidelity of 
our calibrated LL spectrum of HR~6348 is 0.5\% or better.  
Beyond, the spectroscopic fidelity grows worse, with a limit 
of $\sim$1\% at 33~\mum\ and indeterminant beyond 37~\mum.

\section{Assembling a Full Low-Resolution Spectrum of HR~6348} 

To produce a full low-resolution spectrum of HR~6348, we
(1) combined the LL spectrum described here with the SL 
spectrum described by IRS-TR 12002, (2) shifted the LL1
spectrum multiplicatively to match the Red PU photometry
determined by IRS-TR 11002 (Sloan \& Ludovici 2011), and (3)
shifted the remaining orders multiplicatively to eliminate
discontinuities between them.

\begin{figure} 
  \begin{center}
     \epsfig{file=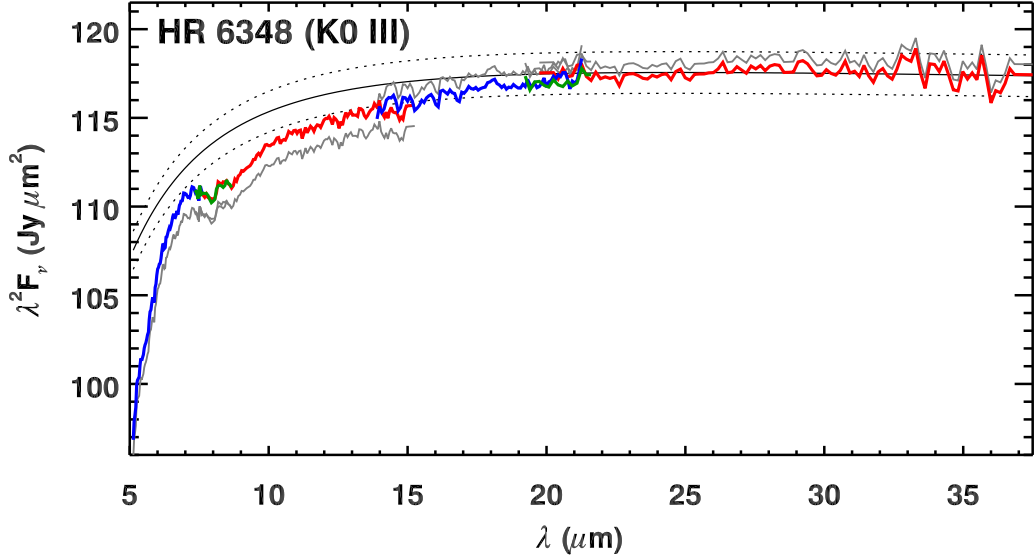, width=13.8cm}
  \end{center}
\caption{
---The low-resolution spectrum of HR~6348.  The data in gray 
are prior to renormalization; the colored data are after.  
Table~2 gives the multiplicative corrections.  A 4400~K 
Engelke function is plotted in black, with the dotted lines
depicting a $\pm$1\% range.}
\end{figure}

\bigskip
\begin{center}
\begin{tabular}{lc} 
\multicolumn{2}{c}{\bf Table 2---Multiplicative corrections for HR 6348} \\
\hline
{\bf Spectral segment} & {\bf Multiplicative correction} \\
\hline
SL2 and bonus & 1.0108 \\
SL1           & 1.0103 \\
LL2 and bonus & 0.9937 \\
LL1           & 0.9950 \\
\hline
\end{tabular}
\end{center}
\bigskip

Figure~5 illustrates the process, and Table 2 lists the 
multiplicative corrections applied to each segment.  The
correction for LL1 is determined by forcing synthetic
photometry from the spectrum to match the actual Red 
PU photometry (IRS-TR 11002).  The normalizations of
SL2 to SL1 and LL2 to LL1 take advantage of the fact that
the bonus orders are observed simultaneously with the 
second-order spectra and overlap the first-order spectra.
The normalization between SL1 and LL2 was determined by
hand to eliminate any discontinuities across the boundary.  
The largest order-to-order correction, between SL1 and LL2,
was 1.7\%, with SL1 having to come up to meet LL2, as is
usually the case due to the pointing-induced loss of flux in 
the smaller SL slit.

We recommend that this spectrum of HR~6348 serve as the 
truth spectrum to calibrate all of the SL data obtained by 
the IRS, and all LL data prior to the change in detector
settings at the beginning of IRS Campaign 45.

\section{HD 173511} 

\begin{figure} 
  \begin{center}
     \epsfig{file=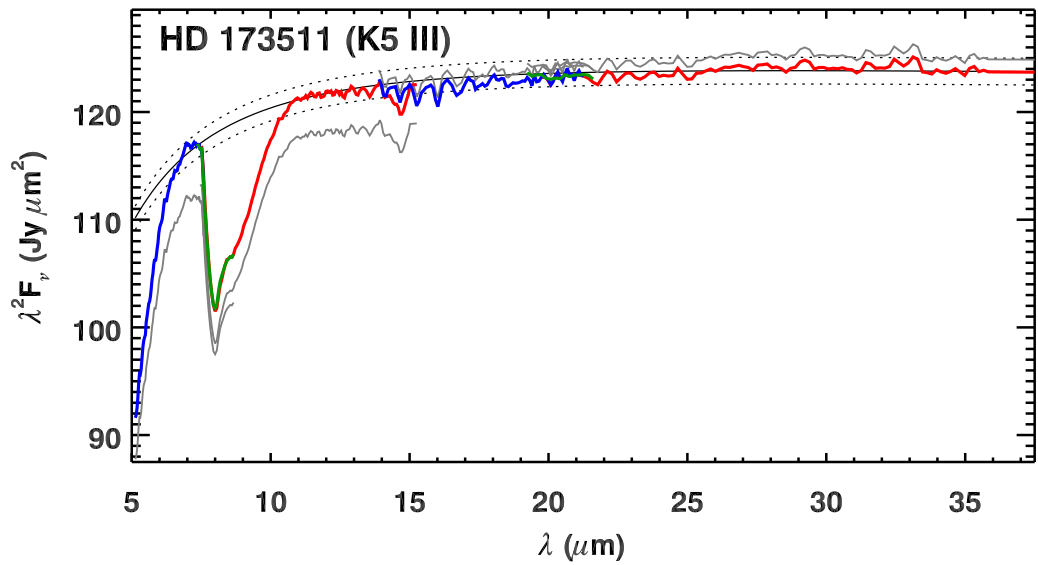, width=13.8cm}
  \end{center}
\caption{
---The low-resolution spectrum of HD~173511.  As in Fig.~5,
gray data depict the spectrum before renormalization and
colored data after.  Table~3 gives the corrections.  A 3800~K
Engelke function is plotted in black, with a $\pm$1\% range
indicated by the dotted lines.}
\end{figure}

The limited number of observations of HR~6348 from Campaign
45 to the end of the cryogenic mission drives the need for a 
second standard to calibrate the LL data after the change in
detector settings.  Using HR~6348 as the standard, we
calibrated all SL observations of HD~173511 and all LL
observations up to Campaign 44 (inclusive).  Following the
usual rejection procedures for the SL data (see IRS-TR 12001,
Sloan \& Ludovici 2012a) and using all of the LL data, we
generated the spectrum shown in gray in Figure~6.  

\bigskip
\bigskip
\begin{center}
\begin{tabular}{lc} 
\multicolumn{2}{c}{\bf Table 3---Multiplicative corrections for HD 173511} \\
\hline
{\bf Spectral segment} & {\bf Multiplicative correction} \\
\hline
SL2 and bonus & 1.0438 \\
SL1           & 1.0304 \\
LL2 and bonus & 0.9931 \\
LL1           & 0.9907 \\
\hline
\end{tabular}
\end{center}
\bigskip

As with HR~6348, we have replaced some data at the ends of
the orders in the spectrum of HD~173511.  We use a 3800 K
Engelke function to replace all data from 35.83~\mum\ to the 
red in LL1.  We replaced the first four elements of LL1 
(19.74--20.25~\mum) with data from LL2, and the two red-most 
elements in LL2 (21.18--21.27~\mum) and the two red-most
elements in the bonus order (21.44-21.61~\mum) with data from 
LL1.

The renormalization of the spectrum follows the procedure for
HR~6348.  To reproduce the Red PU and MIPS 24-\mum\ 
photometry, we shifted LL1 by almost 1\% (see Table 2 in 
IRS-TR 11002).  We then brought the remaining spectral 
segments into alignment.  Table 3 gives the multiplicative 
corrections.  The 3--4\% shifts in SL, while larger than for 
HR~6348, are typical corrections for low-resolution IRS 
spectra.

The resulting spectrum should have a spectroscopic fidelity 
similar to HR~6348 in all orders except LL1.  The OH bands 
are stronger in HD~173511 than in HR~6348, and we have not 
attempted to determine if the observed structure in LL1 is 
due to OH or to fringing.  Despite this uncertainty, the 
deviations from a smooth spectrum in LL1 are smaller than 
$\pm$0.5\% for $\lambda < $ 33~\mum\ and only slightly worse 
past that point.  Beyond 35.8~\mum, we have applied a 3800 K 
Engelke function and do not estimate the fidelity.

We recommend the use of this truth spectrum, in conjunction
with HR~6348, to calibrate LL IRS observations obtained in
Campaign~45 or later (i.e.\ 2007 October 29 and on).

\section{Conclusion} 

This report describes the construction of a truth spectrum 
for the early K giant HR~6348 from IRS observations of it 
and the A dwarfs $\alpha$~Lac and $\delta$~UMi.  This 
spectrum required mitigation for (1) artifacts in the 
vicinity of Pfund-$\alpha$ at 7.45~\mum, (2) a red excess
apparent in the LL spectrum of $\alpha$~Lac, possibly due to
a debris disk, and (3) fringing in LL1 in the spectrum of
HR~6348.  Using HR~6348 as a standard, we also calibrated a
truth spectrum for the late K giant HD~173511.

\bigskip
\begin{center}
\begin{tabular}{lccrccl} 
\multicolumn{7}{c}{\bf Table 4---Intended and actual flux densities} \\
\hline
{\bf Target} & {\bf Intended}       & {\bf Actual}        & {\bf $\Delta$~~} &
               {\bf Intended}       & {\bf Actual}        & {\bf ~~$\Delta$}\\
             & {\bf $F_{22}$ (mJy)} & {\bf $F_{22}$ (mJy)} & {\bf (\%)}     & 
               {\bf $F_{24}$ (mJy)} & {\bf $F_{24}$ (mJy)} & {\bf (\%)}     \\
\hline
HR~6348   & 235.0 & 235.0 &    0.00 & 209.4 & 209.5 & 0.05 \\
HD~173511 & 246.9 & 246.7 & $-$0.08 & 220.0 & 220.2 & 0.09 \\
\hline
\end{tabular}
\end{center}
\bigskip

Table 4 compares the intended flux densities to the actual
results of synthetic photometry for IRS Red Peak-Up and MIPS
at 24~\mum\ for the truth spectra of both K giants.  In the 
22--24~\mum\ range, the differences are less than 0.1\%, as 
designed.  The accuracy of the spectra at other wavelengths 
depends on the accuracy of the shape.  To longer wavelengths, 
the spectra follow an Engelke function to within $\pm$1\% or 
better (Fig.\ 5 and 6).  

To shorter wavelengths, both spectra diverge by more than 1\% 
from the expected continuum shape.  HR~6348 drops more 
steeply to shorter wavelengths.  For HD~173511, if we assume
that the local peak at 7~\mum, between the CO absorption to 
the blue and SiO absorption the red, is slightly below the 
actual continuum in that region (due to overlap of wings of 
the two molecular bands), then this spectrum does not drop
steeply enough, perhaps by $\sim$2\%.  These small errors in
the shape of the SL are a natural consequence of small 
throughput errors from the individual pointings coadded to
generate the plotted spectra, and these throughput errors,
which are a function of both wavelength and pointing, have
been the subject of many previous technical reports (see
IRS-TR 12001 for a recent analysis).  Ultimately, they must
lead to limits on the photometric fidelity (i.e., overall
shape) of the shape of any spectrum observed with SL.

The truth spectra described here are available electronically.
They can be obtained at:
http://isc.astro.cornell.edu/IRS/TruthSpectra.

\end{document}